\documentclass[12pt]{article}
\usepackage{epsfig}
\newcommand{\la}[1]{\label{#1}}
\newcommand{\ur}[1]{(\ref{#1})}

\newcommand{\eq}[1]{eq.~(\ref{#1})}
\newcommand{\eqs}[2]{eqs.~(\ref{#1},\ref{#2})}
\newcommand{\Eq}[1]{Eq.~(\ref{#1})}

\newcommand{\half}{\frac{1}{2}}

\def\Tr{\mbox{Tr}}

\def\beq{\begin{equation}}
\def\eeq{\end{equation}}
\def\bea{\begin{eqnarray}}
\def\eea{\end{eqnarray}}

\begin{document}
\begin{flushright} 
IFUP-TH/2002-18 \\
NORDITA-2002-HE-25
\end{flushright}


\vskip 2true cm
\begin{center}
{\Large\bf Center vortices and $k$-strings} 

\vskip 1.5true cm

{\large\bf Luigi Del Debbio$^*$ and Dmitri Diakonov$^{\diamond}$} \\
\vskip .8true cm
$^*${\it Dipartimento di Fisica and INFN,  Via Buonarroti 2, 
I-56127 Pisa, Italy} \\
\vskip .2true cm
$^\diamond$ {\it NORDITA, Blegdamsvej 17, DK-2100 Copenhagen \O, Denmark} \\
\vskip .2true cm
E-mail: {\small ldd@df.unipi.it, diakonov@nordita.dk}
\end{center}
\vskip 1true cm
\begin{abstract}
\noindent
The vortex contribution to the $k$-string tensions is computed for
SU($N$) gauge theories. We deduce the surface densities needed to
reproduce the sine scaling and the Casimir scaling formulae, recently
obtained from numerical simulations on the lattice. We find that such
densities need to grow linearly in $N$, which in turn suggests that
the vortex scenario can hardly reproduce the physics of confinement at
large $N$.
\end{abstract}
\vskip 1true cm

\section{Introduction}

In a pure glue SU($N$) gauge theory one expects an asymptotic area law
for large Wilson loops in the fundamental representation of the group
and, in general, in all representations with nonzero `$N$-ality'. A
representation is said to have zero $N$-ality if it transforms trivially 
under the center of the group $Z_N$, or, equivalently, if
it is contained in the direct product of some number of adjoint
representations. Wilson loops in $N$-ality-zero representations cannot
exhibit the asymptotic area law since the static source in such
representations can be screened by gluons in the adjoint representation. 

From the confinement point of view, a special role is played by
representations given by rank-$k$ antisymmetric tensors,
with $k=1,\ldots,(N-1)$. The value $k=1$ corresponds to the fundamental
representation whereas $k=N-1$ corresponds to the representation
conjugate to the fundamental. In general, the rank-$(N-p)$
antisymmetric representation is complex conjugate to the rank-$p$
one, and has the same dimension. The dimension of the rank-$k$
antisymmetric representation is
\beq
d(k,N)=C^k_N=\frac{N!}{k!(N-k)!}.
\la{dim}
\eeq
Normalizing the generators in the rank-$k$ representation as
\beq
[T^aT^b]=if^{abc}T^c,\qquad f^{abc}f^{dbc}=\delta^{ad}N,
\la{norm}\eeq
one finds the eigenvalue of the quadratic Casimir operator to be
\beq
T^aT^a=\frac{N+1}{2N}\,k(N-k)\,{\bf 1}. 
\la{Casimirop}\eeq

None of the $k$-representations are contained in the direct product of
adjoint representations, so they have a nonzero $N$-ality.
Furthermore, {\em any} $N$-ality nonzero representation is contained
in a direct product of a certain rank-$k$ antisymmetric representation
and some number of adjoint representations. In other words, {\em all}
representations are equivalent, from the confinement point of view, to
some rank-$k$ antisymmetric representation. (If $k=0$, it is a
$N$-ality-zero representation.) Therefore, all Wilson loops fall into
$N$ classes according to what $k$-representation is the given source
equivalent to, modulo adjoint representations. Correspondingly, there
are only $N$ string tensions $\sigma(k,N),\; k=1...N,$ to characterize
confinement in all possible representations of the trial source, with
$\sigma(N,N)=0$ and $\sigma(1,N)=\sigma_{\rm fund}(N)$. $\sigma(k,N)$
can be interpreted as the tension of a string connecting $k$ quarks to
$k$ antiquarks. Clearly, such a configuration is stable only if the
energy is less than the energy of $k$ fundamental strings connecting
the same color sources. The symmetry under charge conjugation, already
mentioned above, implies that $\sigma(k,N)=\sigma(N-k,N)$, leaving
only $\left[N/2\right]$ independent quantities. Finally, it is
worthwhile to remember that exact factorization in the large-N limit
requires:
\beq
R(k,N)=\frac{\sigma(k,N)}{\sigma(1,N)} \stackrel{N \rightarrow \infty}
\longrightarrow k.
\la{asR}\eeq

In this work, we examine the vortex contribution to rank-$k$
antisymmetric Wilson loops, which leads to a prediction for
$\sigma(k,N)$ as a function of the density of $l$-vortices
carrying flux $l$. By comparing this prediction with recent numerical 
results for the $k$-string tensions~\cite{LT,ldd01}, we are able to actually
compute the vortex densities that are needed to fit the results from Monte
Carlo simulations.

\section{Wilson loops in $k$-representations}

Let the Wilson loop in the fundamental representation be
some unitary matrix $U^\alpha_\beta$. In $k=2$ antisymmetric
representation the same Wilson loop is then
$U^{[\alpha_1\alpha_2]}_{[\beta_1\beta_2]}
=\half\left(U^{\alpha_1}_{\beta_1}U^{\alpha_2}_{\beta_2}-
U^{\alpha_1}_{\beta_2}U^{\alpha_2}_{\beta_1}\right)$. For a general
rank-$k$ representation one has to antisymmetrize the product
of $k$ fundamental unitary matrices $U^\alpha_\beta$. We introduce
the trace of the Wilson loop $W(k,N)$ in the $k$-representation,
normalized in such a way that all $W$'s are unity at $U={\bf 1}$.
We have:
\bea
\nonumber
W(1,N)&=&\frac{1}{N}\,\Tr\,U,\\
\nonumber
W(2,N)&=&\frac{2}{N(N-1)}\left((\Tr\,U)^2-\Tr\,U^2\right),\\
\la{Wk}
W(3,N)&=&\frac{1}{d(3,N)}\left((\Tr\,U)^3-2\Tr\,U^2\,\Tr\,U
+3\Tr\,U^3 \right),
\eea
and so on. 

We now turn to the center vortices; there are also $(N-1)$ types of
them and we shall name them $l$-vortices, $l=1,\ldots,(N-1)$.  By
definition of an $l$-vortex, the Wilson loop in the fundamental
representation of the SU($N$) gauge group, winding around the
$l$-vortex in the transverse plane is gauge-equivalent to a diagonal
unitary matrix being a non-trivial element of the group center, 
as the radius of the Wilson loop tends to infinity:
\beq
U^\alpha_\beta={\rm P}\exp i \oint A_\mu dx^\mu
\rightarrow  \left( \begin{array} {ccc}
   z_N^l & & \\
   & \ddots & \\
   & & z_N^l
\end{array}\right)\in Z(N),\qquad z_N^l\equiv e^{2\pi i l /N}.
\la{W1}
\eeq
From \Eq{Wk} we immediately find the corresponding traces of Wilson
loops in the $k$-representation:
\beq
W(k,N)=z_N^{kl}=e^{2\pi i k l/N}.
\la{Wk1}
\eeq
\section{Area law for $k$-loops from $l$-vortices}

Let us consider the following simplest scenario: The Yang--Mills
vacuum is populated by random center $l$-vortices. Let us assume
the vortices are statistically independent. The probablility that
$n_l$ vortices pierce a given Wilson loop is then given by the
Poisson distribution,
\beq
P_{n_l}=\frac{\bar n_l^{n_l}}{n_l!}e^{-\bar n_l},\qquad
\sum_{n_l} P_{n_l}=1,
\la{P}
\eeq
where the average number of $l$-vortices going through a surface
spanned over the Wilson loop is
\beq
\bar n_l=\rho(l,N)\cdot {\rm Area}
\la{barn}
\eeq
where $\rho(l,N)$ is the average density of $l$-vortices in the
SU($N$) gauge theory, piercing any two-dimensional plane; its
dimension is $1/{\rm cm}^2$.

According to \eq{Wk1} each $l$-vortex contributes to the $k$-loop
a factor $z_N^{kl}$. Assuming the Poisson distribution of the
vortices \ur{P} we get for the average $k$-loop:

\bea
\nonumber
W(k,N)&=&\prod_{l=1}^{N-1}\sum_{n_l=0}^\infty\,
P_{n_l} \left(z_N^{kl}\right)^{n_l} \\
\nonumber
&=& \prod_{l=1}^{N-1}\exp(-\bar n_l+\bar n_l\,z_N^{kl})\\
\la{Wk2}
&=& \exp\left[-{\rm Area}\,
\sum_{l=1}^{N-1}\rho(l,N)(1-z_N^{kl})\right].
\eea
This is the area law with the string tension
\beq
\sigma(k,N)=\sum_{l=1}^{N-1}\rho(l,N)\left(1-e^{2\pi i k l/N}\right).
\la{s1}
\eeq
Using the fact that $\rho(l,N)=\rho(N-l,N)$, the above equation can be
rewritten as
\beq
\sigma(k,N)=\sum_{l=1}^{\left[N/2\right]}
	2 \rho(l,N) \left(1-\cos \frac{2\pi k l}{N}\right)
\la{s1bis}
\eeq
if $N$ is odd, while for even $N$ one gets:
\beq
\sigma(k,N)=\sum_{l=1}^{N/2-1}
	2 \rho(l,N) \left(1-\cos \frac{2\pi k l}{N}\right) +
	\rho\left(\frac{N}{2},N\right)\left(1 - (-1)^k\right).
\la{s1ter}
\eeq
In both cases, one obtains explicitely a positive, real number for the
string tension. In particular, for the SU(2) gauge group there is
only one type of center vortex with $l=1$ and $z_2=-1$, and only
one non-trivial representation with $k=1$, the fundamental one. In
this case we recover the well-known result:
\beq
\sigma(1,2)=2\rho(1,2).
\la{s2}
\eeq
In SU(3) one can have $k,l=1,2$; however $k=2$ is the antitriplet
which is conjugate to the fundamental representation, while $l=2$
corresponds to the vortices conjugate to those with $l=1$ so that
their densities must be equal, $\rho(1,3)=\rho(2,3)$. In this case we
get from \eq{s1} only one non-trivial string tension:
\beq 
\sigma(1,3)=\sigma(2,3)=3\rho(1,3)=3\rho(2,3).  
\la{s3}
\eeq
Starting from SU(4) there are different types of $l$-vortices and
correspondingly different string tensions for $k$-loops. One could
naively assume that the vortex density is independent of $l$,
i.e. $\rho(l,N)=\rho_N$. This oversimplified scenario can be
immediately ruled out by noting that it leads to $k$-string tensions
that are independent of $k$ and proportional to $N$, namely
$\sigma(k,N)=\rho_N N$, for each $k$. The inconsistency of this result
with large-$N$ factorization immediately points toward more refined
scenarios. The next Section is devoted to the determination of the
vortex densities needed to reproduce the $k$-string tensions recently
computed by numerical simulations.

\section{Two scenarios}

Two types of behaviour of the $k$-strings have been discussed in the
literature: the ``sine'' regime~\cite{DS-95,HSZ-98,Witten-97},
\beq
\sigma_{\rm sin}(k,N)=\sigma_1\,\frac{\sin\frac{\pi k}{N}}{\sin\frac{\pi}{N}},
\la{Sin}
\eeq
and the ``Casimir'' regime~\cite{AOP-84},
\beq
\sigma_{\rm cas}(k,N)=\sigma_1\,\frac{k(N-k)}{N-1}.
\la{Cas}
\eeq
Recent numerical simulations show that the ``sine'' scaling hypothesis
provides an accurate description of the $k$-string spectrum to the
level of a few percent~\cite{ldd01}. On the other hand, even the
Casimir scaling formula turns out to be a good approximation at the
level of 10\%~\cite{LT,ldd01}. In both cases $\sigma_1$ is the string
tension in the fundamental representation of the SU($N$) group; it may
depend somewhat on $N$ but asymptotically it is believed that
$\sigma_1=O(N^0)$, i.e. that it is independent of $N$. The two
formulae are plotted in Fig. 1.

\begin{figure}[t]
\begin{center}
\epsfig{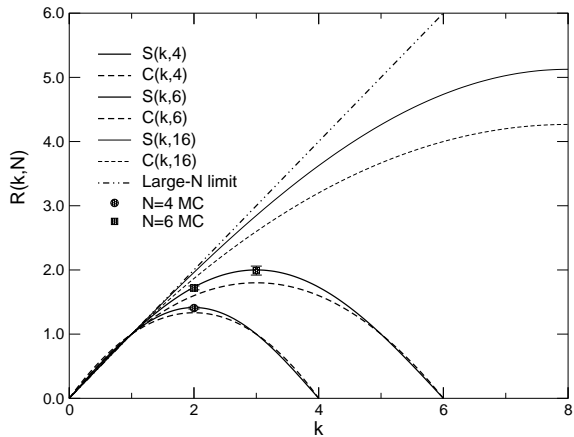}
\end{center}
\vspace{-.5cm}\caption{{\small Ratios of $k$-string tensions to the
fundamental ones for gauge groups SU(4), SU(6), SU(16) and SU($\infty$),
in the ``sine'' and ``Casimir'' scenarios. Also shown are lattice-measured
values for SU(4) and SU(6)~\cite{ldd01}.}}
\end{figure}

The two regimes for the string tension correspond to two different
types of behaviour for the two-dimensional center vortices densities
$\rho(l,N)$, as functions of their `flux number' $l$. To find
$\rho(l,N)$, one needs to equate the general expression \ur{s1} to the
appropriate string tension, \eq{Sin} or \eq{Cas}. \Eq{s1} is in fact a
Fourier series, and it is not difficult to calculate $\rho(l,N)$ from
the inverse Fourier transformation. We find:
\bea 
\la{rhoS} 
\rho_{\rm sin}(l,N)&=&\sigma_1\,\frac{1}{N}\, 
\frac{1}{\cos\frac{\pi}{N}-\cos\frac{2\pi l}{N}},\\ 
\la{rhoC} 
\rho_{\rm cas}(l,N)&=& \sigma_1\,\frac{1}{2(N-1)}\, 
\frac{1}{\sin^2\frac{\pi l}{N}}.  
\eea 
The two formulae are plotted in Fig. 2. In both cases one reproduces the
relation $\rho(l,N)=\rho(N-l,N)$, as it should be since $(N-l)$-vortices are
complex conjugate to the $l$-vortices. The asymptotics of \eqs{rhoS}{rhoC} at
large $N$ and fixed $l$ are: 
\bea
\la{rhoSa}
\rho_{\rm sin}(l,N)&\rightarrow &
\sigma_1\left[\frac{2N}{\pi^2(4l^2-1)}
+\frac{1}{6N}\frac{4l^2+1}{4l^2-1}+O\left(\frac{1}{N^2}\right)\right],\\
\nonumber \\ \la{rhoCa}
\rho_{\rm cas}(l,N)&\rightarrow &
\sigma_1\left[\frac{N^2}{2(N-1)\,\pi^2\,l^2}
+\frac{1}{6(N-1)}+O\left(\frac{1}{N^2}\right)\right]. 
\eea
Since the string tension in the fundamental representation
$\sigma_1=O(N^0)$, we see that both regimes above require the
two-dimensional densities of vortices to grow linearly with $N$. At
the same time, the transverse sizes of vortices are stable in $N$, as
it follows from the dimensional analysis of ref.  \cite{DM}. It means
that at large $N$ center vortices are inevitably strongly overlapping,
and the notion of vortices looses sense.

In principle, one can imagine a regime where the density of vortices
does not grow with $N$. For example, one can assume, as a matter of 
an exercise, that only maximal-flux $l$-vortices exist, with 
$l\approx \half N$. Apart from being a rather unnatural hypothesis,
it is in contradiction with the $k$-string tensions measured in lattice 
simulations \cite{LT,ldd01}.

\begin{figure}[t]
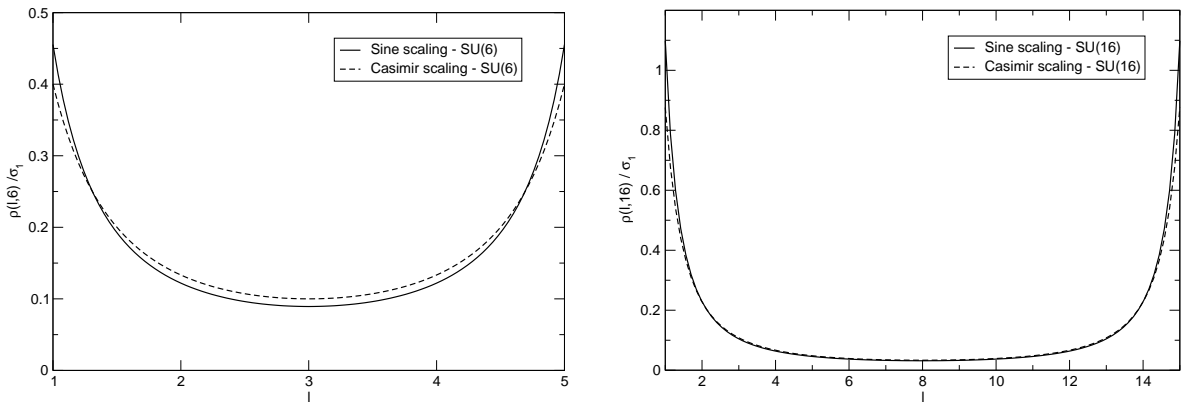

\begin{center}
\epsfig{file=rho_v_su6.eps,width=7.5cm}
\hspace{.5cm}\epsfig{file=rho_v_su16.eps,width=7.5cm}
\end{center}
\vspace{-.5cm}
\caption{{\small Densities of center vortices assuming the ``sine''
(solid) or ``Casimir'' (dashed) scenarios for the $k$-string tensions.
The gauge groups are SU(6) (left) and SU(16) (right). In
both scenarios, the vortex densities increase and become peaked around
small values of $l$ as the number of colors grows large.}}
\end{figure}

\section{Conclusions}
  
From the viewpoint of the confinement behaviour, sources in various
representations of the SU($N$) gauge group fall into $N$ classes with
the same string tension $\sigma(k,N)$ for all representations
belonging to the same $k$-class. The value $k=0$ (or equivalently $k=N$
since $\sigma(k,N)=\sigma(N-k,N)\,$) labels $N$-ality zero
representations {\it i.e.} adjoint and those which arise in a direct
product of any number of adjoint representations. The
string tension here is zero as sources in such representations 
are screened by gluons. Sources in $N$-ality nonzero representations can be
partially screened: the representation belongs to the class of the rank-$k$
antisymmetric tensor if it is found in the direct product of that
representation with some number of adjoint ones. Therefore, the $k$-string
tensions $\sigma(k,N)$ are the only fundamental ones which can appear in the
asymptotic area law. The dynamics of confinement is encoded in these numbers. 

Assuming that confinement is driven by center vortices populating the
Yang--Mills vacuum, one derives relations between the surface densities
of $l$-vortices piercing any given two-dimensional plane, where
$l=1,\ldots,(N-1)$ is the flux of an SU($N$) center vortex. In fact,
these densities $\rho(l,N)$ are Fourier coefficients of the $k$-string
tensions $\sigma(k,N)$.  We have found the densities needed to
reproduce two popular regimes for $k$-strings: the ``sine'' and the
``Casimir'' regimes, see Figs. 1,2.  In both cases we find that the
density of vortices should rise linearly with $N$. It means that the
vortex scenario of confinement can hardly be stable in $N$: at large
enough $N$ vortices have to overlap, and the vortex language probably
becomes senseless.
  
\bigskip
\noindent
{\bf Acknowledgements} \\
LDD thanks the Niels Bohr Institute for
hospitality in the early stages of this work.


\begin{thebibliography}{99}
%
\bibitem{LT}
B. Lucini and M. Teper, {\it Phys. Lett.} {\bf B501} (2001) 128; {\it Phys.
Rev.} {\bf D64} (2001) 105019; hep-lat/0110004. 
%
\bibitem{ldd01}
L. Del Debbio, H. Panagopoulos, P. Rossi and E. Vicari, hep-th/0111090.
%
\bibitem{DS-95}
M. R. Douglas and S. H. Shenker,
Nucl. Phys. B {\bf 447}, 271 (1995).
%
\bibitem{HSZ-98}
A. Hanany, M. J. Strassler, and A. Zaffaroni,
Nucl. Phys. B {\bf 513}, 87 (1998).
%
\bibitem{Witten-97}
E. Witten, Nucl. Phys. B {\bf 500}, 3 (1997); B {\bf 507}, 658 (1997).
%
\bibitem{AOP-84}
J. Ambjorn, P. Olesen, and C. Peterson,
Nucl. Phys. B {\bf 240}, 533 (1984).
%
\bibitem{DM}
D. Diakonov and M. Maul, hep-lat/0204012.  
%
%
\end{thebibliography}
\end{document}